\documentclass{article}

\usepackage{PRIMEarxiv}

\usepackage[utf8]{inputenc} 
\usepackage[T1]{fontenc}    
\usepackage{hyperref}       
\usepackage{url}            
\usepackage{booktabs}       
\usepackage{amsfonts}       
\usepackage{nicefrac}       
\usepackage{microtype}      
\usepackage{lipsum}
\usepackage{fancyhdr}       
\usepackage{graphicx}       
\graphicspath{{media/}}     
\usepackage{graphicx}      
\usepackage{natbib}        
\setcitestyle{numbers,square}
\usepackage{amsmath,amssymb,amsfonts}

\newtheorem{definition}{Definition}
\newtheorem{lemma}{Lemma}
\newtheorem{theorem}{Theorem}
\newtheorem{proposition}{Proposition}
\newtheorem{remark}{Remark}
\newtheorem{assumption}{Assumption}
\newtheorem{problem}{Problem}

\title{Bearing-Based Target Entrapping Control of Multiple Uncertain Agents With Arbitrary Maneuvers
}

\author{
  Haifan Su, Ziwen Yang, Shanying Zhu, Cailian Chen, Wenbin Yu\\
  Department of Automation, Shanghai Jiao Tong University, Shanghai 200240, China \\
  Key Laboratory of System Control and Information Processing, Ministry of Education of China \\
  Shanghai Engineering Research Center of Intelligent Control and Management, Shanghai 200240, China\\
  \texttt{\{sjtu-suhaifan,1106385445,shyzhu,cailianchen,yuwenbin\}@sjtu.edu.cn} \\
}

\begin{document}
\Large
\normalsize
\maketitle
\large

\begin{abstract}
\normalsize
This paper is concerned with bearing-based cooperative target entrapping control of multiple uncertain agents with arbitrary maneuvers including shape deformation, rotations, scalings, etc. A leader-follower structure is used, where the leaders move with the predesigned trajectories, and the followers are steered by an estimation-based control method, integrating a distance estimator using bearing measurements and a stress matrix-based formation controller. The signum functions are used to compensate for the uncertainties so that the agents' accelerations can be piecewise continuous and bounded to track the desired dynamics. With proper design of the leaders' trajectories and a geometric configuration, an affine matrix is determined so that the persistently exciting conditions of the inter-agent relative bearings  can be satisfied since the bearing rates are related to different weighted combinations of the affine matrix vectors. The asymptotic convergence of the estimation error and control error is proved using Filipov properties and cascaded system theories. A sufficient condition for inter-agent collision avoidance is also proposed. Finally, simulation results are given to validate the effectiveness of the method in both 2D and 3D cases.
\end{abstract}


\section{Introduction}
Cooperative target entrapping control, which aims to steer the agents to entrap a target and reach desired distances or displacements with respect to the target, has attracted much attention in recent years \cite{yu_cooperative_2021}. For entrapping control problems, great challenges arise from the uncertainty caused by disturbances or unknown dynamics. To solve such a problem, the disturbance observer-based controllers \citep{yu_cooperative_2021}, adaptive approximation-based methods, and robust controllers with signum functions are developed \citep{ju_mpc-based_2022}.
Note that the above works assume that the displacements among the agents and the target or global positions of the target are available. In severe environments such as underwater surroundings, however, the sensing abilities of the agents are restricted and such position information is usually unknown. 

For sensing-restricted formation control problems, a preferable way is to design control methods based on bearing measurements, which impose minimal requirements on agents' sensing abilities. 
Furthermore, to solve the bearing based target entrapping control problems, estimation-based control schemes are usually used for the agents to localize and surround the target 
simultaneously. For instance, position estimators \citep{zou_distributed_2021} and displacement estimators \citep{dou_target_2020,hu_bearing-only_2022} using bearing measurements together with estimation-based controllers are proposed for localizing and enclosing the target with a circular orbit. 
The convergence of the estimation errors and control errors is guaranteed by the so-called persistent excitation conditions. To be specific, the agents need to measure the bearing information of a target or a neighbor from different angles for localizing it and further entrapping control \citep{yang_distributed_2021}. However, it should be pointed out that the above methods in \citep{zou_distributed_2021,dou_target_2020,hu_bearing-only_2022} only steer the agents to entrap the target with a constant angular velocity and hence the persistent excitation conditions are easy to satisfy. Besides, the unknown uncertainties are not considered. To adapt to the practical environmental changes, the agents are required to cooperatively entrap the target with more general trajectories, i.e., achieving an entrapping formation with any maneuvers including scalings, rotations, shape deformations, etc. 
The uncertainties in the agents' dynamics make it more challenging to achieve flexible and sophisticated  maneuvers in sensing-restricted environment, which requires new design of robust bearing-based entrapping controllers for the agents.

To achieve bearing-based formation control with uncertainties, bearing rigidity-based methods are often used. The work \citep{Lixiaolei-auv} proposes an adaptive model uncertainty estimation-based formation control method. \citep{bae_distributed_2022} proposes sufficient conditions on the disturbance bounds which guarantee the ultimate boundedness of the control errors under several bearing rigidity-based controllers. \citep{robust-bearing} proposes a signum function-based controller to tackle the bounded uncertainties and the formation errors are asymptotically convergent. However,  \citep{Lixiaolei-auv,bae_distributed_2022,robust-bearing} can only achieve fixed formation shapes due to the inflexibility of the bearing rigidity structure. 

To achieve entrapping formation with scaling and rotation maneuvers, a bearing-based displacement estimator and a complex Laplacian-based control law are proposed in \citep{yang_distributed_2021}. Furthermore, the works \citep{tang_formation_2021,tang_relaxed_2022} propose bearing-based controllers to achieve time-varying formation control without estimators. The errors converge if the desired relative bearings between each couple of neighboring agents are persistently exciting. However, the control methods in \citep{yang_distributed_2021,tang_formation_2021,tang_relaxed_2022}  can not compensate for the uncertainties, which may result in the unknown and even nonsmooth dynamics of the agents and hence the persistently exciting conditions are hard to satisfy. Moreover, the geometric formation patterns in \citep{yang_distributed_2021,tang_formation_2021,tang_relaxed_2022} are still fixed. In these cases, the angular velocities of all the agents related to the formation center is consistent. To achieve more flexible maneuvers such as shape deformation, the agents' angular velocities may be different and they can not be described by a common rotation matrix in $\mathcal{SO}(2)$ or $\mathcal{SO}(3)$ as in \citep{yang_distributed_2021,tang_formation_2021,tang_relaxed_2022}. Therefore, the dynamics of the inter-agent bearing rates determined by the angular velocities are harder to describe, which even increases the challenges of satisfying the persistently exciting conditions. The affine formation control method \citep{zhao_affine_2018}, which contains formation maneuvers in the positive and negative edge weights of a stress matrix and a constant geometric configuration, can be used to design displacement-based formation control laws for achieving arbitrary maneuvers \citep{zhao_affine_2018}. However, the method is hard to be extended to bearing-based cases since the relationship between the neighboring agents is merely defined on the displacements but not the bearings.

\par Motivated by the above observation, this paper aims to achieve robust target entrapping control with arbitrary maneuvers using bearing measurements. 
The main contributions are summarized as follows.
\textbf{i)} An estimation-based control method, consisting of a bearing-based estimator to estimate the distance in real time and a stress matrix-based formation controller based on the distance estimations, is proposed for the followers. The signum functions are used in the controller to compensate for nonsmooth uncertainties so that the agents' accelerations can be piecewise continuous and bounded to track the desired dynamics. Compared with \citep{zou_distributed_2021,dou_target_2020,hu_bearing-only_2022,yang_distributed_2021,tang_formation_2021,tang_relaxed_2022}, the agents can track arbitrary entrapping formation maneuvers such as shape deformations, scalings, and rotations.  
\textbf{ii)} By fully exploiting the relations between the inter-agent bearing rates and the different weighted combinations of the affine matrix vectors, we characterize the sufficient conditions on the leaders' trajectories and the geometric configuration, so that the persistently exciting conditions can be satisfied by the resulting affine matrix with the proposed estimation based control method.
Then the asymptotic convergence of both the estimation errors and control errors are proved using Filipov properties and cascaded system theories.

\textit{Notations:} For a vector $x_i=[x_{i1},...,x_{id}]^\mathrm{T}\in\mathbb{R}^d$, one has col$(x_i):=[x_1^\mathrm{T},...,x_n^\mathrm{T}]^\mathrm{T}\in\mathbb{R}^{dn}$, $\|x\|=\sqrt{x^\mathrm{T}x}$ and $\|x\|_1=\sum_{i=1}^n\sum_{j=1}^d|x_{ij}|$, where $|\cdot|$ is the absolute value. $\text{sgn}(x_i)=[\text{sgn}(x_{i1}),...,\text{sgn}(x_{id})]^\mathrm{T}$ is the signum function, where $\text{sgn}(x_{ij})=1$ if $x_{ij}>0$, $\text{sgn}(x_{ij})=-1$ if $x_{ij}<0$, and $\text{sgn}(x_{ij})=0$ if $x_{ij}=0$ for all $j=1,...,d$. $O_{d}$ and $I_d$ are the null matrix and identity matrix with $d$ dimensions, respectively. $\textbf{0}_d$ and $\textbf{1}_d$ are the vectors of all zeros and all ones, respectively. $\otimes$ is the Kronecker product. $\lambda_{\min}(\cdot)$ and $\lambda_{\max}(\cdot)$ denote the minimum and the largest eigenvalues of the matrix, respectively.

\section{Problem Formulation}
Consider a moving target and a multi-agent system consisting of $n_l$ leaders labeled by $\mathcal{V}_l=\{1,...,n_l\}$ and $n_f$ followers labeled by $\mathcal{V}_f=\{n_l+1,...,n\}$. The target has the following dynamics
\begin{equation}\label{target}
	\dot{p}_0=v_0,
\end{equation}
where $p_0$ and $v_0$ are its position and velocity, respectively. Each leader agent moves with the desired dynamics
\begin{equation}\label{leader i}
	\dot{p}_i=v_i^*, \forall i\in\mathcal{V}_l,
\end{equation}
where $v_i^*$ is the desired velocity of leader $i$.
Each follower agent $i$ has the following dynamics:
\begin{equation}\label{dot eta i}
	\left\{
	\begin{array}{lr}
		\dot{p}_i=v_i,\\
		\dot{v}_i=u_i+\Delta_i, \forall i\in\mathcal{V}_f,
	\end{array}
	\right.
\end{equation} 
where  $p_i$, $v_i$, and $u_i\in\mathbb{R}^d(d\geq2)$ denote the position, velocity, and acceleration control input, respectively. $\Delta_i(t)\in\mathbb{R}^d$ denotes the uncertainty which satisfies $\|\Delta_i(t)\|\leq\bar{\Delta}, \forall t\geq0$, where $\bar{\Delta}$ is a bounded constant. 

\begin{figure}[htpb]
	\centering
	\includegraphics[scale=1]{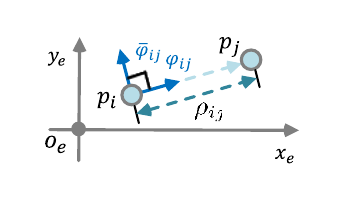}
	\caption{Notations in 2D cases. 
	}
	\label{2D_notation}
\end{figure}

In the multi-agent system, define a graph as $\mathcal{G}=\{\mathcal{V},\mathcal{E}\}$, where $\mathcal{V}=\mathcal{V}_l\cup\mathcal{V}_f$ denotes the node set and $\mathcal{E}=\{(i,j),i\in\mathcal{V},j\in\mathcal{N}_i\}$ denotes the edge set. Here, each vertex $i$ denotes an agent. 
$\mathcal{N}_i:=\{j\in \mathcal{V}:(i,j)\in \mathcal{E}\}$ is the neighbor set of agent $i$. $(i,j)$ denotes the edge from vertex $i$ to $j$, which means agent $i$ can sense the relative bearings
$$\varphi_{ij}=\frac{p_j-p_i}{\|p_j-p_i\|}=\frac{p_{ij}}{\rho_{ij}}$$ in a global reference frame as in Fig.\ref{2D_notation} and receive the velocity information $v_j$ from agent $j$. 
$p_{ij}$ and $\rho_{ij}$ denote the displacement and distance from agent $i$ to $j$, respectively. In 2D cases, $\bar{\varphi}_{ij}$ is obtained by rotating $\varphi_{ij}$ ninety degrees anticlockwise. 
The edges are undirected, i.e., $(i,j)\in\mathcal{E}\Leftrightarrow(j,i)\in\mathcal{E}$. 

We introduce a stress matrix ${L}\in\mathbb{R}^{n\times n}$ associated with $\mathcal{G}$, where its $(i,j)$th off-diagonal element is $-l_{ij}$ if $(i,j)\in\mathcal{E}$, and $0$ otherwise, and the $(i,i)$th diagonal element is $\sum_{j\in\mathcal{N}_i}l_{ij}$. 
$l_{ij}>0$ and $l_{ij}<0$ represent an attracting force and repelling force in edge $(i,j)\in\mathcal{E}$, respectively. For more details about the stress matrix, please refer to \cite{zhao_affine_2018}. 
The stress matrix $L$ is normally a positive semi-definite matrix that can be partitioned to 
\vskip-1em
\begin{equation}\label{bearing laplacian}
	{L}=\begin{bmatrix}{L}_{ll}&{L}_{lf}\\{L}_{fl}&{L}_{ff}\end{bmatrix},
\end{equation}
based on the partition of leaders and followers, 
where ${L}_{ll}\in\mathbb{R}^{n_l\times n_l}$, ${L}_{ff}\in\mathbb{R}^{n_f\times n_f}$.

The entrapping formation to be achieved is defined as follows.
\begin{definition}[Entrapping Formation]\label{bearing entrapping formation}
	An entrapping  formation is defined to satisfy the following conditions
	\begin{enumerate}
		\item $p_0(t)=\frac{1}{n}\sum_{i\in\mathcal{V}}p_{i}^*(t),$
		\item $p_i^*(t)=A(t)r_i+b(t),\ \forall\ i\in\mathcal{V},$
	\end{enumerate}
	where $p_i^*(t)$ is the desired position of each agent. 
	$r=[r_l^\mathrm{T},r_f^\mathrm{T}]^\mathrm{T}=[\text{col}(r_i)_{i\in\mathcal{V}_l}^\mathrm{T},\text{col}(r_i)_{i\in\mathcal{V}_f}^\mathrm{T}]^\mathrm{T}$ denotes a constant configuration. $\bar{r}=\sum_{i\in\mathcal{V}}r_i/n$ denotes the center of the configuration $r$. $A(t)$ and $b(t)$ denote the transformation matrix and the position of the whole formation  which are determined by 
	\begin{align}
		A(t)=&\Big(\sum_{i\in\mathcal{V}_l}p_i^*(t)\tilde{r}_i^\mathrm{T}\Big)\Big(\sum_{i\in\mathcal{V}_l}\tilde{r}_i\tilde{r}_i^\mathrm{T}\Big)^{-1},\label{A}\\
		b(t)=&\frac{1}{n_l}\sum_{i\in\mathcal{V}_l}p_{i}^*(t)-A(t)\bar{r}_l,\label{b}
	\end{align}
	where $\bar{r}_l=\sum_{i\in\mathcal{V}_l}r_i/n_l$ and $\tilde{r}_i=r_i-\bar{r}_l$ denote the centroid of the leaders' configuration $r_l$ and the offset of each leader to the centroid, respectively. $\{r_i\}_{i\in\mathcal{V}_l}$ affinely span $\mathbb{R}^{d}$ so that $(\sum_{i\in\mathcal{V}_l}\tilde{r}_i\tilde{r}_i^\mathrm{T})^{-1}$ exists.
\end{definition}

\begin{figure}[htpb]
	\centering
	\includegraphics[scale=0.2]{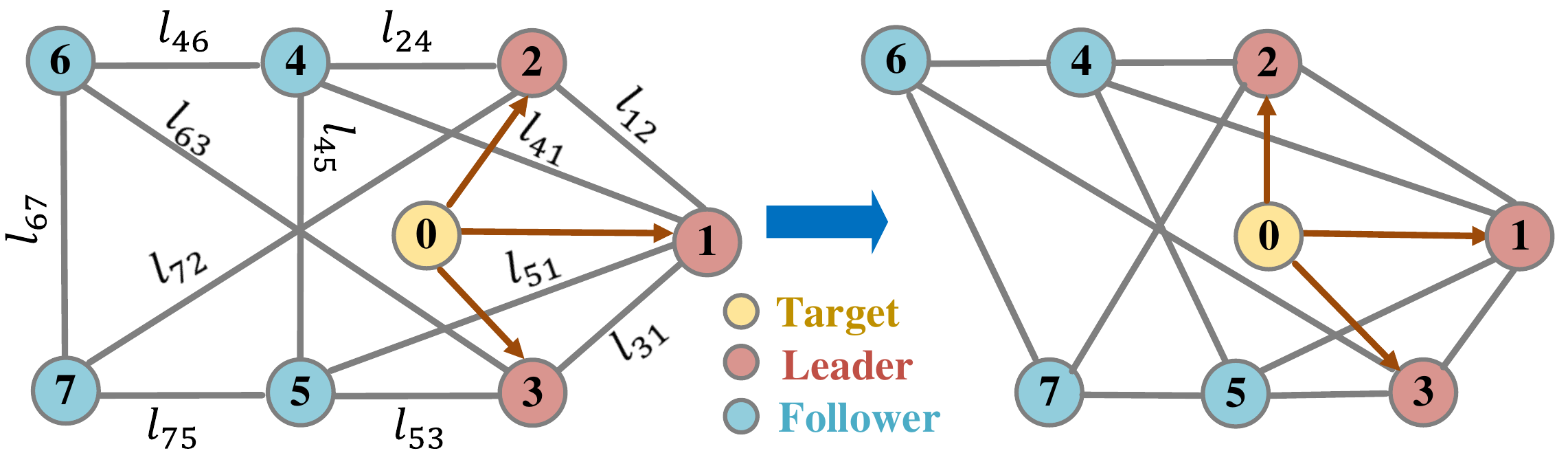}
	\caption{Examples of the entrapping formation. 
	}
	\label{topology}
\end{figure}

The first condition in \emph{Definition \ref{bearing entrapping formation}} describes that the center of the entrapping formation is consistent with the target. The second condition is a time-varying affine transformation of the constant configuration $r$, which is inspired by the stress matrix-based work \citep{zhao_affine_2018}. When $A(t)$ has the form as $A(t)=\alpha(t)N(t)$, where $\alpha(t)\in\mathbb{R}$ and $N(t)\in\mathcal{SO}(2)$ or $\mathcal{SO}(3)$ denote the scaling and rotation maneuvers, the entrapping formation is similar to that in \citep{yang_distributed_2021,tang_formation_2021,tang_relaxed_2022}. However, aside from scalings and rotations, our formation also contains shape deformation maneuvers as in Fig.\ref{topology}. 

\begin{remark}
	Examples of the entrapping formation in 2D are shown in Fig.\ref{topology}. The red arrows from the target to the leaders denote that only the leaders can observe the target. The target and the constant configuration are set as $p_0=[\frac{2 }{3},0]^\mathrm{T}$ and $r=[2,0,1,1,1,-1,0,1,0,-1,-1,1,-1,-1]^\mathrm{T}$, respectively. In the left entrapping formation, we consider that $p^*=r$. In the right entrapping formation, Leader 2 and Leader 3 move in opposite directions to $p_2^*=[\frac{2}{3},1]^\mathrm{T}$ and $p_3^*=[\frac{4}{3},-1]^\mathrm{T}$, respectively, and hence the entrapping formation becomes $p^*=[2,0,\frac{2}{3},1,\frac{4}{3},-1,-\frac{1}{3},1,\frac{1}{3},-1,-\frac{4}{3},1,-\frac{2}{3},-1]^\mathrm{T}$. It can be easily proved that the two conditions in \emph{Definition \ref{bearing entrapping formation}} are satisfied in both of the formations in Fig.\ref{topology}.
\end{remark}

According to \eqref{A} and \eqref{b}, the followers' desired positions can be uniquely determined by the leaders' desired positions $p_l^*$ and constant configuration $r$. Therefore, the first condition in \emph{Definition \ref{bearing entrapping formation}} is equivalent to
\begin{align}
	p_0(t)=&\frac{1}{n}\sum_{i\in\mathcal{V}}p_i^*(t)=A(t)\bar{r}+b(t)\nonumber\\
	=&\frac{1}{n_l}\sum_{i\in\mathcal{V}_l}p_{i}^*(t)+A(t)(\bar{r}-\bar{r}_l),\label{center}
\end{align}
which can be achieved by designing $p_l^*$ and $r$ only. 
Since we consider that the leaders move with the desired dynamics, i.e., $p_l=p_l^*$, the remaining objective is to achieve $p_f\to p_f^*$, where
\begin{equation}\label{leader follower 1}
	p_f^*=[I_{n_f}\otimes A(t)]r_f+I_{n_f}\otimes b(t).
\end{equation}

According to \eqref{A} and \eqref{b}, $A(t)$ and $b(t)$ are time-varying global variables which are determined by all the leaders positions in $p_l^*$ and configuration $r_l$. To achieve $p_f\to p_f^*$ distributedly, we adopt the following assumption on the stress matrix $L$.

\begin{assumption}\label{L ass}
	$L$ satisfies the following two conditions:
	\begin{enumerate}
		\item $(L\otimes I_d)r=0$.
		\item $L_{ff}$ is positive definite and $L_{fl}\neq0$.
	\end{enumerate}
\end{assumption}
The condition $(L\otimes I_d)r=0$ in \emph{Assumption \ref{L ass}} can be written in a distributed form as $\sum_{j\in\mathcal{N}_i}l_{ij}(r_i-r_j)=0$. The desired position $p_i^*,i\in\mathcal{V}$ of each agent has the following relationship with its neighbors' desired positions $p_j^*,j\in\mathcal{N}_i$:
\begin{align}
		\sum_{j\in\mathcal{N}_i}l_{ij}(p_i^*-p_j^*)&=\sum_{j\in\mathcal{N}_i}l_{ij}(Ar_i+b-Ar_j-b)\nonumber\\
		&=A\sum_{j\in\mathcal{N}_i}l_{ij}(r_i-r_j)=0,\label{omega distributed}
\end{align}
which can be used to design the distributed control law in the following section.
\eqref{omega distributed} can also be written as the matrix form $\bar{L}_{ff}p_f^*+\bar{L}_{fl}p_l^*=0$, which yields 
\begin{equation}\label{relationship}
	p_f^*=-\bar{L}_{ff}^{-1}\bar{L}_{fl}p_l^*,
\end{equation}
under the second condition of \emph{Assumption \ref{L ass}},
where $\bar{L}_{ff}=L_{ff}\otimes I_d$ and $\bar{L}_{fl}=L_{fl}\otimes I_d$.  \eqref{relationship} also means that $p_f^*$ can be uniquely determined by the matrix $L$ and the leaders' desired positions $p_l^*$.

To sum up, the problem to be solved is formulated as follows.
\begin{problem}\label{problem 1}
	Consider the multi-agent system consisting of $n_l$ leaders and $n_f$ followers with graph $\mathcal{G}$, where the leader agents are moving with the desired trajectories and the followers can only observe the relative bearings $\varphi_{ij}$ and obtain the relative velocities $v_{ij}$ with their neighbors. Design the desired trajectories of the leaders and a control method for the followers so that the entrapping formation in \emph{Definition \ref{bearing entrapping formation}} is achieved.
\end{problem}

\section{Main Methods}
In this section, we aim to design a method for solving \emph{Problem \ref{problem 1}} in 2D. The desired trajectories of the leaders are first designed. Then a bearing-based control method, which consists of a distance estimator using bearing measurements and a robust estimation-based controller,  is designed for the followers. 

\subsection{Desired trajectories for the leaders}
Some sufficient conditions are firstly proposed for the leaders' desired trajectories $p_i^*,i\in\mathcal{V}_l$ to guarantee that the entrapping formation in \emph{Definition \ref{bearing entrapping formation}} is admissible. 


Denote the time derivative of the maneuver matrix by $\dot{A}(t)=[a_1(t),a_2(t)]$, where $a_1$ and $a_2$ are the two columns of $\dot{A}$. For each couple of neighboring agents in $(i,j)\in\mathcal{E}$, the desired relative velocity is denoted by
\begin{align}
	v_{ij}^*(t)&=\dot{A}(t)r_j+\dot{b}(t)-\dot{A}(t)r_i-\dot{b}(t)\nonumber\\
	&=r_{ij1}a_1(t)+r_{ij2}a_2(t),\label{linear combine}
\end{align}
which is a weighted linear combination of the affine matrix vectors. $r_{ij1}$ and $r_{ij2}$ are the two constants in $r_{ij}$. 
Denote $v_{ij}^*(t)=z_{ij}^*(t)\gamma_{ij}^*(t)$, where $z_{ij}^*=\|v_{ij}^*\|$ and $\gamma_{ij}^*$ is varying rate of the relative velocity orientations satisfying $\gamma_{ij}^*=v_{ij}^*/\|v_{ij}^*\|$ if $\|v_{ij}^*\|\neq0$, $\gamma_{ij}^*=0$ otherwise. 
Then the following conditions are adopted. 


\textit{Condition 1:}  
$\dot{p}_{i}^*$ and $\ddot{p}_{i}^*$ are bounded for all $i\in\mathcal{V}_l,j\in\mathcal{N}_i$. 

\textit{Condition 2:} There exist $\epsilon_{v}>0$ and $\sigma_{v}>0$, such that 
\begin{align}
	&\int_t^{t+\sigma_{v}}\|\dot{\gamma}_{ij}^*(\tau)\| d\tau>\epsilon_{v}, \forall (i,j)\in\mathcal{E}, t\geq0	\label{leader PE2}.
\end{align}

\textit{Condition 1} is adopted so that the leaders' desired positions and velocities are smooth. According to \eqref{relationship} and \emph{Definition \ref{bearing entrapping formation}}, it also guarantees that the followers' desired dynamics are smooth and ${\gamma}_{ij}^*$ is differential. 
\textit{Condition 2} is related to the desired bearing rate $\dot{\varphi}_{ij}^*=(I-{\varphi}_{ij}^*{\varphi}_{ij}^{*\mathrm{T}})\gamma_{ij}^*$. 
The two conditions can be naturally satisfied by designing the leaders' desired positions and the geometric configuration. 



\subsection{Distance estimator using bearing measurements}
To achieve the entrapping formation in \emph{Definition \ref{bearing entrapping formation}}, we need to achieve $p_i\to p_i^*$ for all $i\in\mathcal{V}_f$. Since the position  $p_i$ of each follower is unknown but bearing measurements are available, a bearing-based estimator is proposed for each follower to estimate the relative distance $\forall i\in\mathcal{V}_f, j\in\mathcal{N}_i$: 
\begin{equation}\label{follower estimator}
	\begin{aligned}
		\dot{\hat{\rho}}_{ij}=\varphi_{ij}^\mathrm{T}v_{ij}\!-\!k_1(\hat{\rho}_{ij}|\bar{\varphi}_{ij}^\mathrm{T}\dot{\varphi}_{ij}|\!-\!|\bar{\varphi}_{ij}^\mathrm{T}v_{ij}|),
	\end{aligned}
\end{equation}
where $\hat{\rho}_{ij}$ and $v_{ij}=v_j-v_i$ denote the distance estimation and relative velocities from agent $i$ to $j$, and $k_1>0$ is an estimator gain.   

The convergence of the estimation errors $\tilde{\rho}_{ij}:=\hat{\rho}_{ij}-\rho_{ij}$ for all $i\in\mathcal{V}_f,j\in\mathcal{N}_i$ is shown in the following.
\begin{lemma}\label{est converge}
	Under the estimator \eqref{follower estimator} and assuming that there is no collision between each couple of agents, one has
	\begin{enumerate}
		\item $\lim_{t\to\infty}\tilde{\rho}_{ij}=\bar{\rho}_{ij}$, where $\bar{\rho}_{ij}$ is some constant;
		\item $\bar{\rho}_{ij}=0$ if there exist positive reals $\epsilon_\omega$ and $\sigma_\omega$ such that 
		\begin{equation}\label{PE condition}
			\int_{t}^{t+\sigma_\omega}|\bar{\varphi}_{ij}^\mathrm{T}(\tau)\dot{\varphi}_{ij}(\tau)|d\tau>\epsilon_\omega;
		\end{equation}
		\item If $\bar{\rho}_{ij}\neq0$ for some $i\in\mathcal{V}_f,j\in\mathcal{N}_i$, then $\varphi_{ij}$ converges to some unit constant.
	\end{enumerate}
\end{lemma}
The proof can be referred to \citep{cao_safe_2021} and hence omitted here.

Note that \eqref{PE condition} is the so-called persistently exciting condition, which is a sufficient condition to ensure the convergence of the estimation error $\tilde{\rho}_{ij}$ for all $i\in\mathcal{V}_f,j\in\mathcal{N}_i$.

\subsection{Estimation-based controller}
In this part, a robust controller is proposed for each follower to achieve the entrapping formation while compensating for the uncertainty $\Delta_i$.
Based on the distance estimation  $\hat{\rho}_{ij}$, one can estimate the inter-agent displacement by $\hat{p}_{ij}=\hat{\rho}_{ij}\varphi_{ij},\forall i\in\mathcal{V}_f,j\in\mathcal{N}_i$. The controller is proposed as 
\begin{equation}\label{follower controller}
	\begin{aligned}
		u_i\!=&\sum_{j\in\mathcal{N}_i}l_{ij}(k_p\hat{p}_{ij}+k_vv_{ij})+k_{\Delta}\text{sgn}\Bigl[\sum_{j\in\mathcal{N}_i}l_{ij}(\hat{p}_{ij}+v_{ij})\Bigr],
	\end{aligned}
\end{equation}
for all $i\in\mathcal{V}_f$,
where 
$\sum_{j\in\mathcal{N}_i}l_{ij}(k_p\hat{p}_{ij}+k_vv_{ij})$ is inspired by the affine relation \eqref{relationship},  and $k_\Delta\text{sgn}(\cdot)$ is used to compensate for the uncertainty $\Delta_i$.
$k_{p}$, $k_{v}$, and $k_{\Delta}$ are all positive and constant controller gains. 

\section{Stability Analysis}

Denote the control errors as $\delta_{p_i}:=p_{i}-p_{i}^*$ and $\delta_{v_i}:=v_{i}-v_{i}^*$ for all $i\in\mathcal{V}$. Stack them into vectors as 
$\delta_{p_f}=\text{col}(\delta_{p_i})_{i\in\mathcal{V}_f}$ and $\delta_{v_f}=\text{col}(\delta_{v_i})_{i\in\mathcal{V}_f}$.  
Then one has
\begin{align}
	\dot{e}_f=B_{f}(e_f,t)+\tilde{p}_f,
	\label{perturbed follower}
\end{align}
where
\begin{align}
	B_f\!=\!-\!\underbrace{\begin{bmatrix}
			O_{2n_f}&I_{2n_f}\\k_p\bar{L}_{ff}&\!\!k_v\bar{L}_{ff}
	\end{bmatrix}}_A\!\!e_f\!-\!
	\!\begin{bmatrix}
		\textbf{0}_{2n_f}\\\!k_{\Delta}\text{sgn}[\bar{L}_{ff}(\delta_{p_f}\!\!+\!\delta_{v_f})]\!-\!\Delta_f\!+\!\dot{v}_f^*
	\end{bmatrix}\label{B_f}
\end{align}
is the nominal part and
\begin{align}
	\tilde{p}_f\!=\!\begin{bmatrix}\textbf{0}_{2n_f}\\\text{col}\Big\{\!\!\sum\limits_{j\in\mathcal{N}_i}[k_pl_{ij}\tilde{\rho}_{ij}\varphi_{ij}\!+\!k_\Delta\text{sgn}(l_{ij}\tilde{\rho}_{ij}\varphi_{ij}) ]\Big\}_{i\in\mathcal{V}_f}\end{bmatrix}\label{tilde p_f}
\end{align}
denotes the perturbation which depends on the estimation errors $\tilde{\rho}_{ij},\forall i\in\mathcal{V}_f,j\in\mathcal{N}_i$, where $\dot{v}_f^*=\text{col}(\dot{v}_i^*)_{i\in\mathcal{V}_f}$ and $\Delta_f:=\text{col}(\Delta_i)_{i\in\mathcal{V}_f}$ contain the desired accelerations and uncertainties of all the followers.

The stability of the nominal error system $\dot{e}_f=B_f(e_f,t)$ is first given.


\begin{theorem}\label{nominal stability}
	Under the controller \eqref{follower controller} and \emph{Assumption} \textit{\ref{L ass}}, the nominal system 
	$\dot{e}_f=B_f(e_f,t)$ is exponentially stable if 
	\begin{equation}\label{gains}
		\left\{
		\begin{array}{lcl}
			k_p>0\\k_vI_{2n_f}>\bar{L}_{ff}^{-1}\\k_{\Delta}>\bar{\Delta}+\|\dot{v}_f^*\|
		\end{array}
		\right.,
	\end{equation}
	is satisfied and no inter-agent collision happens.
\end{theorem}

	According to \citep{zhao_affine_2018}, the followers' desired accelerations satisfy $\dot{v}_f^*=-\bar{L}_{ff}^{-1}\bar{L}_{fl}\dot{v}_l^*$. According to \emph{Assumption \ref{L ass}}, $\bar{L}_{ff}$ is a positive definite constant matrix and $\bar{L}_{fl}\neq0$. Hence, $\|\dot{v}_f^*\|$ is bounded if $\|\dot{v}_l^*\|$ is bounded. Recall that $\bar{\Delta}$ is a bounded constant. Then $k_p$, $k_v$, and $k_{\Delta}$ can be chosen as bounded constants.
	
	Consider the Lyapunov function
	\begin{equation}
		V_c=\frac{1}{2}e_f^\mathrm{T}\!\!\begin{bmatrix}
			(k_p\!+\!k_v)\bar{L}_{ff}^2&\bar{L}_{ff}\\\bar{L}_{ff}&\bar{L}_{ff}
		\end{bmatrix}\!e_f\!:=\frac{1}{2}e_f^\mathrm{T}Pe_f.
	\end{equation}
	Since $\bar{L}_{ff}>0$ and $k_vI_{2n_f}>\bar{L}_{ff}^{-1}$, one has $(k_p+k_v)\bar{L}_{ff}^2-\bar{L}_{ff}\bar{L}_{ff}^{-1}\bar{L}_{ff}=(k_p+k_v)\bar{L}_{ff}^2-\bar{L}_{ff}>0.$ According to the Schur complement, one has $P>0$ and $V_c$ is positive definite and $\lambda_{\max}(P)>\lambda_{\min}(P)>0$.
	Since the controller \eqref{follower controller} is piecewise continuous, the solution of the system is obtained in Filipov sense, i.e., $\dot{V}_c\in^{a.e.}\dot{\tilde{V}}_c=\bigcap_{\xi\in\partial V}\xi^\mathrm{T}K[B_{f}(e_f,t)]$, where $a.e.$ means almost everywhere and $K[B_{f}(e_f,t)]:=\bigcap_{\theta>0}\bigcap_{\mu \zeta=0}\overline{co}\ g(D(e_f,\theta)-\zeta,t).$ $\mu$, $\overline{co}$, and $D(e_f,\theta)$ denote the Lebesgue measure, the convex closure, and the open ball of radius $\theta$ centered at $e_f$, respectively. 
	Then one has
	\begin{align}
		\dot{\tilde{V}}_c=&(k_p\!+\!k_v)\delta_{p_f}^\mathrm{T}\bar{L}_{ff}^2\delta_{v_f}+\delta_{v_f}^\mathrm{T}\bar{L}_{ff}{\delta}_{v_f}+({\delta}_{p_f}^\mathrm{T}+{\delta}_{v_f}^\mathrm{T})\bar{L}_{ff}\dot{\delta}_{v_f}\nonumber\\
		=&-k_p\delta_{p_f}^\mathrm{T}\bar{L}_{ff}^2\delta_{p_f}-\delta_{v_f}^\mathrm{T}(k_v\bar{L}_{ff}^2-\bar{L}_{ff})\delta_{v_f}+(\delta_{p_f}^\mathrm{T}+\nonumber\\
		&\delta_{v_f}^\mathrm{T})\bar{L}_{ff}\big\{-k_\Delta K\big\{\text{sgn}[\bar{L}_{ff}(\delta_{p_f}+\delta_{v_f})]\big\}+\Delta_f\!-\!\dot{v}_f^*\big\}\nonumber\\
		\leq&-k_p\delta_{p_f}^\mathrm{T}\bar{L}_{ff}^2\delta_{p_f}-\delta_{v_f}^\mathrm{T}(k_v\bar{L}_{ff}^2-\bar{L}_{ff})\delta_{v_f}\nonumber\\
		&-(k_\Delta-\bar{\Delta}-\|\dot{v}_f^*\|) \|\bar{L}_{ff}(\delta_{p_f}+\delta_{v_f})\|_1
	\end{align}
	where the fact $x\cdot K[\text{sgn}(x)]=\{|x|\}$, $\forall x\in\mathbb{R}$, is used. Thus, $\dot{\tilde{V}}_c$ is a singleton and $\dot{V}_c=\dot{\tilde{V}}_c$ from \citep{robust-bearing}. Since 
	$k_{\Delta}>\bar{\Delta}+\|\dot{v}_f^*\|$, one has
	\vskip-1.5em
	\begin{align}
		\dot{V}_c\leq&-k_p\delta_{p_f}^\mathrm{T}\bar{L}_{ff}^2\delta_{p_f}-\delta_{v_f}^\mathrm{T}(k_v\bar{L}_{ff}^2-\bar{L}_{ff})\delta_{v_f}\nonumber\\
		=&-e_f\begin{bmatrix}
			k_p\bar{L}_{ff}^2&O_{2n_f}\\O_{2n_f}&k_v\bar{L}_{ff}^2-\bar{L}_{ff}
		\end{bmatrix}:=-e_f^\mathrm{T}Qe_f.
	\end{align}
	Then one has  $\dot{V}_c\leq[-2\lambda_{\min}(Q)/\lambda_{\max}(P)]V_c,$
	which leads to
	\begin{align}\label{ef bound}
		\|e_f(t)\|\leq\sqrt{\frac{\lambda_{\max}(P)}{\lambda_{\min}(P)}}\|e_f(0)\|e^{-\frac{\lambda_{\min}(Q)}{\lambda_{\max}(P)}t}.
	\end{align}
	Since $k_p\bar{L}_{ff}>0$ and $k_vI_{2n_f}>\bar{L}_{ff}^{-1}$, one has $k_p\bar{L}_{ff}^2>0$ and $k_v\bar{L}_{ff}^2-\bar{L}_{ff}>0$. According to the Schur complement, $Q$ is positive definite, and hence $\lambda_{\min}(Q)>0$.  
	The system $\dot{e}_f=B_f(e_f,t)$ is exponentially stable. The proof is completed.

%

The convergence of the perturbation $\tilde{p}_f$ has been shown in \emph{Lemma \ref{est converge}}. Then we will show the convergence of the error $e_f$ in the perturbed system \eqref{perturbed follower}. 
\begin{theorem}\label{perturbed convergence}
	Under the estimator \eqref{follower estimator} and the controller \eqref{follower controller}, the entrapping formation in \emph{Definition \ref{bearing entrapping formation}} is asymptotically achieved if the leaders' trajectories satisfy \textit{Conditions 1-2}, \emph{Assumption} \textit{\ref{L ass}} is satisfied, the gains satisfy \eqref{gains}, and no inter-agent collision happens.
\end{theorem}

	The asymptotic stability of the perturbed system \eqref{perturbed follower} can be directly derived from the cascade system theories \cite[Lemma 4.7]{nonlinear} after we prove the asymptotic convergence of $\tilde{p}_f$ and $e_f$ of the nominal system $\dot{e}_f=B_{f}(e_f,t)$. The asymptotic stability of the system $\dot{e}_f=B_{f}(e_f,t)$ is obtained by \emph{Theorem \ref{nominal stability}}. The rest is to prove $\tilde{p}_f\to0$ by checking whether the persistently exciting condition \eqref{PE condition} is satisfied. 
	
	
	According to \cite{yang_distributed_2021}, if \eqref{PE condition} is not satisfied, one has $\lim_{t\to\infty}\bar{\varphi}_{ij}^\mathrm{T}\dot{\varphi}_{ij}=0$. Then we show it is contradicted with \eqref{leader PE2} in \textit{Condition 2} about the leaders' trajectories.  
	According to \emph{Theorem \ref{nominal stability}}, the time derivative of $V_c$ along the perturbed system \eqref{perturbed follower} satisfies 
	\begin{align}
		\dot{V}_c=&-e_f^\mathrm{T}Qe_f+\frac{1}{2}e_f^\mathrm{T}P\tilde{p}_f\nonumber\\
		\leq&-\|e_f\|(\lambda_{\min}(Q)\|e_f\|-\frac{\lambda_{\max}(P)}{2}\|\tilde{p}_f\|)\leq0,
	\end{align}
	if $\|e_f\|\geq\frac{\lambda_{\max}(P)\|\tilde{p}_f\|}{2\lambda_{\min}(Q)}$. According to \emph{Lemma \ref{est converge}},
	$\tilde{p}_f$ converges to some constant $\bar{p}_f$ as $t\to\infty$.
	Similar to \cite[Th. 3]{yang_distributed_2021}, $\delta_{p_f}$ keeps bounded and $\lim_{t\to\infty}\delta_{v_f}=0$.
	For any $(i,j)\in\mathcal{E}$, one has $\rho_{ij}$ being bounded and
	\begin{equation}\label{omega to 0}
		\lim_{t\to\infty}\bar{\varphi}_{ij}^\mathrm{T}\dot{\varphi}_{ij}=\lim_{t\to\infty}\bar{\varphi}_{ij}^\mathrm{T}v_{ij}^*/\rho_{ij}=0,
	\end{equation} 
	which leads to $\lim_{t\to\infty}\bar{\varphi}_{ij}^\mathrm{T}v_{ij}^*=0$. According to Subsection 3.1, one has $v_{ij}^*=z_{ij}^*(t)\gamma_{ij}^*(t)$. 
	Since $z_{ij}^*=\|v_{ij}^*\|$ is bounded, one has
	\begin{align}
		\lim_{t\to\infty}\bar{\varphi}_{ij}^\mathrm{T}\gamma_{ij}^*=0.\label{inequality 1}
	\end{align} 
	According to \emph{Lemma \ref{est converge}}, $\varphi_{ij}$ converges to some constant if $\lim_{t\to\infty}\tilde{\rho}_{ij}\neq0$. In this case, \eqref{inequality 1} yields that ${\gamma}_{ij}^*$ converges to some constant and $\lim_{t\to\infty}\dot{\gamma}_{ij}^*=0$. That is contradicted with \eqref{leader PE2} in \textit{Condition 2}. 
	The same result can be obtained for all $i\in\mathcal{V}_f,j\in\mathcal{N}_i$. Therefore, \eqref{PE condition} in \emph{Lemma \ref{est converge}} is satisfied and one has $\lim_{t\to\infty}\tilde{\rho}_{ij}=0$, i.e., $\lim_{t\to\infty}\tilde{p}_{f}=0$. Then the asymptotic convergence of the control errors $\delta_{p_f}$ and $\delta_{v_f}$ within the perturbed system \eqref{perturbed follower} is obtained by \cite[Lemma 4.7]{nonlinear}. This means that the entrapping formation in \emph{Definition \ref{bearing entrapping formation}} is achieved and the proof is completed.

Note that \emph{Lemma \ref{est converge}}, \emph{Theorem \ref{nominal stability}}, and \emph{Theorem \ref{perturbed convergence}} rely on the assumption of collision avoidance. To drop this assumption, a sufficient condition
on the initial control error $e_f(0)$ and the estimation error $\tilde{\rho}_{ij}(0),i\in\mathcal{V}_f,j\in\mathcal{N}_i$ is given. 
\begin{proposition}\label{collision avoidance}
	Under \emph{Assumption} \textit{\ref{L ass}}, if the initial errors $e_f(0)$ and $\tilde{\rho}_{ij}(0),\forall i\in\mathcal{V}_f,j\in\mathcal{N}_i$ are sufficiently small so that
	\begin{equation}\label{avoidance condition}
		\begin{aligned}
			&\max\Biggl(\sqrt{\frac{\lambda_{\max}(P)}{\lambda_{\min}(P)}}\|e_f(0)\|,\frac{\lambda_{\max}(P)c_e}{2\lambda_{\min}(Q)}\Biggr)\\
			&\leq\Biggl(\!\min_{t\geq0,i,j\in\mathcal{V}}\|p_i^*(t)\!-\!p_j^*(t)\|\!-\!c_c\!\Biggr),
		\end{aligned}
	\end{equation}
	where $c_{e}=\sum_{i\in\mathcal{V}_f}\sum_{j\in\mathcal{N}_i}[k_pl_{ij}|\tilde{\rho}_{ij}(0)|+2k_\Delta ]$ and $c_c$ is chosen so that $\!\min_{t\geq0,i,j\in\mathcal{V}}\|p_i^*(t)\!-\!p_j^*(t)\|\!>c_c>0$, then $\|p_i(t)-p_j(t)\|\geq c_c$ for all $i,j\in\mathcal{V},t\geq0$ and the entrapping formation in \emph{Definition \ref{bearing entrapping formation}} is asymptotically achieved.
\end{proposition}

	According to \emph{Theorem \ref{perturbed convergence}}, one has
	$V_{ij}(t)\leq V_{ij}(0),\forall t\geq0$, i.e., $|\tilde{\rho}_{ij}(t)|\leq|\tilde{\rho}_{ij}(0)|$ for all $i\in\mathcal{V}_f,j\in\mathcal{N}_i$. Therefore, one has
	\begin{align}
		\|\tilde{p}_f(t)\|
		\leq&\sum_{i\in\mathcal{V}_f}\sum\limits_{j\in\mathcal{N}_i}[k_pl_{ij}|\tilde{\rho}_{ij}(0)|+2k_\Delta ]\!\!=c_{e}.\label{ce}
	\end{align}
	According to \eqref{ce} and \emph{Theorem \ref{perturbed convergence}} and, one has
	\begin{align}
		\dot{V}_c
		\leq-\|e_f\|(\lambda_{\min}(Q)\|e_f\|-\frac{\lambda_{\max}(P)c_e}{2}).
	\end{align}
	Then one has $\dot{V}_c\leq0$ for all $\|e_f\|\geq\frac{\lambda_{\max}(P)c_e}{2\lambda_{\min}(Q)}$. 
	According to \cite[Th.4.18]{nonlinear} and \emph{Lemma \ref{nominal stability}}, there exists a positive constant $c_T$ so that 
	\begin{align}
		\|\delta_{p_f}(t)\|\leq\|e_f(t)\|
		\leq\sqrt{\frac{\lambda_{\max}(P)}{\lambda_{\min}(P)}}\|e_f(0)\|,\label{delta pf1} 
	\end{align}
	for all $0\leq t\leq c_T$ and
	\begin{equation}\label{delta pf2}
		\!\|\delta_{p_f}(t)\|\!\leq\!\|e_f(t)\|\!\leq\!\frac{\lambda_{\max}(P)c_e}{2\lambda_{\min}(Q)}
	\end{equation}
	for all $t\geq c_T$.  Combining \eqref{delta pf1} with \eqref{delta pf2} gives
	\begin{equation}\label{delta pf}
		\begin{aligned}
			\|\delta_{p_f}(t)\|\leq&\max\Biggl(\sqrt{\frac{\lambda_{\max}(P)}{\lambda_{\min}(P)}}\|e_f(0)\|,\frac{\lambda_{\max}(P)c_e}{2\lambda_{\min}(Q)}\Biggr), \forall t\geq0.
		\end{aligned}
	\end{equation}
	For all $i\in\mathcal{V}_f,j\in\mathcal{N}_i,t\geq0$, one has
	\begin{align}
		&\|p_i(t)\!-\!p_j(t)\|\nonumber\\
		\!\geq&\|p_i^*(t)\!-\!p_j^*(t)\|\!-\!\|\delta_{p_i}(t)\|\!-\!\|\delta_{p_j}(t)\|\nonumber\\
		\!\geq&\|p_i^*(t)\!-\!p_j^*(t)\|\!-\!\|\delta_{p_f}(t)\|\nonumber\\
		\!\geq&\|p_i^*(t)\!-\!p_j^*(t)\|\!-\!\max\Biggl(\sqrt{\frac{\lambda_{\max}(P)}{\lambda_{\min}(P)}}\|e_f(0)\|,\frac{\lambda_{\max}(P)c_e}{2\lambda_{\min}(Q)}\Biggr)\nonumber\\
		\geq&\|p_i^*(t)-p_j^*(t)\|\!-\!\!\min_{t\geq0,i,j\in\mathcal{V}}\|p_i^*(t)\!-\!p_j^*(t)\|\!+\!c_c\!\geq c_c\label{collision 2}.
	\end{align}
	Moreover, for all $i,j\in\mathcal{V}_l$, one has $\|p_i(t)-p_j(t)\|=\|p_i^*(t)-p_j^*(t)\|$. Therefore, \eqref{collision 2} is true for all $i,j\in\mathcal{V}$,
	which means the inter-agent collision can be avoided. Similar to \emph{Theorem \ref{perturbed convergence}}, we can still obtain that the control error $e_f$ within the perturbed system \eqref{perturbed follower} is asymptotically convergent. The proof is completed.

By \emph{Proposition \ref{collision avoidance}}, collision avoidance can be guaranteed
by adjusting the initial positions of the agents and initial
estimations of the inter-agent distances.

\section{numerical examples}
\subsection{Target entrapping control in 2D}
A multi-agent system with $3$ leaders and $5$ followers is taken into account. The topology structure is shown in Fig.\ref{topology}. The stress matrix ${L}$ is set with $l_{12}=l_{21}=l_{31}=l_{13}=l_{46}=l_{64}=l_{75}=l_{57}=0.2714$, $l_{41}=l_{14}=l_{51}=l_{15}=l_{63}=l_{36}=l_{72}=l_{27}=-0.137$, $l_{24}=l_{42}=l_{53}=l_{35}=0.548$, $l_{45}=l_{54}=0.0685$, and $l_{67}=l_{76}=0.137$, where ${L}_{ff}$ is positive definite and ${L}_{fl}\neq0$, i.e., \emph{Assumption \ref{L ass}} is satisfied. The nominal configuration is set as $r=[2,0,1,1,1,-1,0,1,0,-1,-1,1,-1,-1]^\mathrm{T}$ to satisfy \emph{Assumption \ref{L ass}}. The dynamics of the target are set as $p_0(0)=[\frac{2}{3},0]^\mathrm{T}m$ and $v_0=[0.5,0.5]^\mathrm{T}m/s$. To make the formation center consistent with the target and \textit{Conditions 1}-\textit{2} be satisfied, the desired trajectories of the leaders are set as $p_{01}^*(0)=[4,0]^\mathrm{T}m$, $v_{01}^*(t)=\frac{\pi}{5}\bar{p}_{01}^*(t)m/s$, $p_{02}^*(0)=[2,2]^\mathrm{T}m$, $v_{02}^*(t)=[\pi\|p_{01}^*(t)\|/(5\|p_{02}^*(t)\|)]\bar{p}_{02}^*(t)m/s$,
$p_{03}^*(0)=[2,-2]^\mathrm{T}m$, and $v_{03}^*(t)=[\pi\|p_{01}^*(t)\|/(5\|p_{03}^*(t)\|)]\bar{p}_{03}^*(t)m/s$, where $p_{0i}^*=p_i^*-p_0$, $v_{0i}^*=v_i^*-v_0$, and $\bar{p}_{0i}^*=Rp_{0i}^*$, for $i=1,2,3,$ where $R=[0,-1;1,0]$. Then one has
$\|\dot{v}_f^*\|\leq3.5$. The uncertainty is set as $\Delta_i=0.2\sin(0.1t)$ for each follower. The gains are chosen as $k_1=1$, $k_p=1$, $k_v=1.5$, and $k_\Delta=4$. The other initial states are chosen to satisfy \eqref{avoidance condition}.

\begin{figure}[htpb]
	\centering
	\includegraphics[scale=0.51]{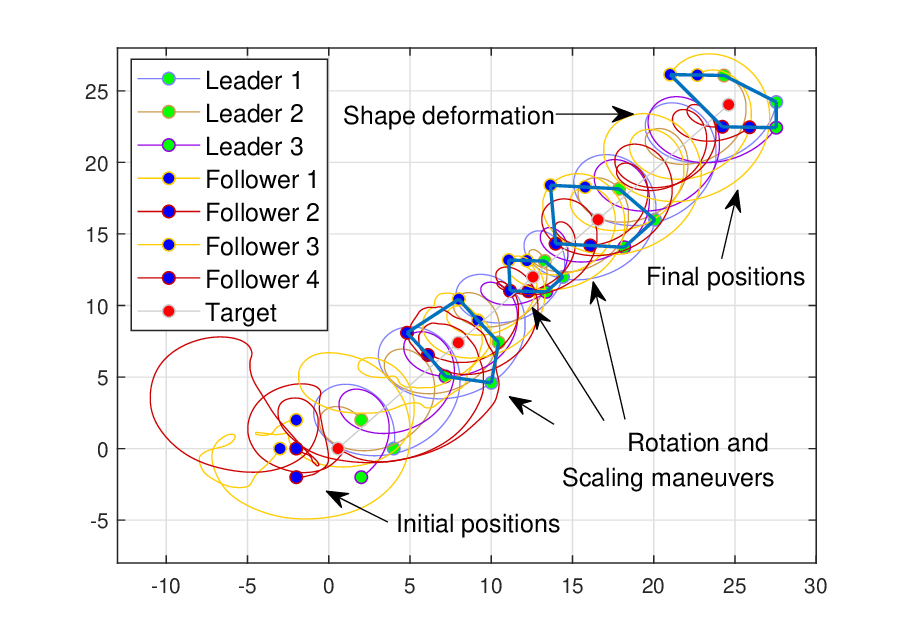}
	\caption{Trajectories of the agents and the target in 2D.}
	\label{trajectory}
\end{figure}
\begin{figure}[htpb]
	\centering
	\includegraphics[scale=0.51]{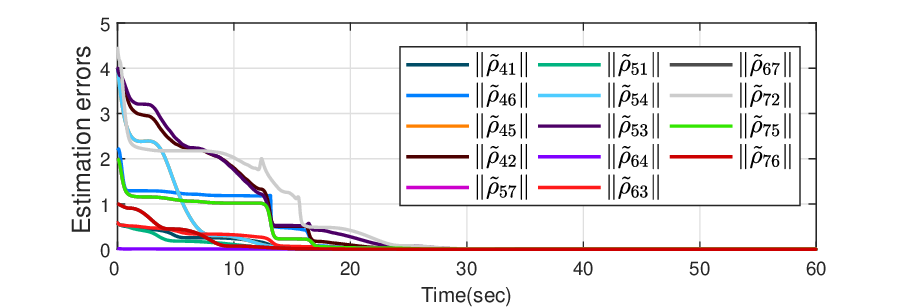}
	\caption{Distance estimation errors in 2D.}
	\label{est}
\end{figure}
\begin{figure}[htpb]
	\centering
	\includegraphics[scale=0.51]{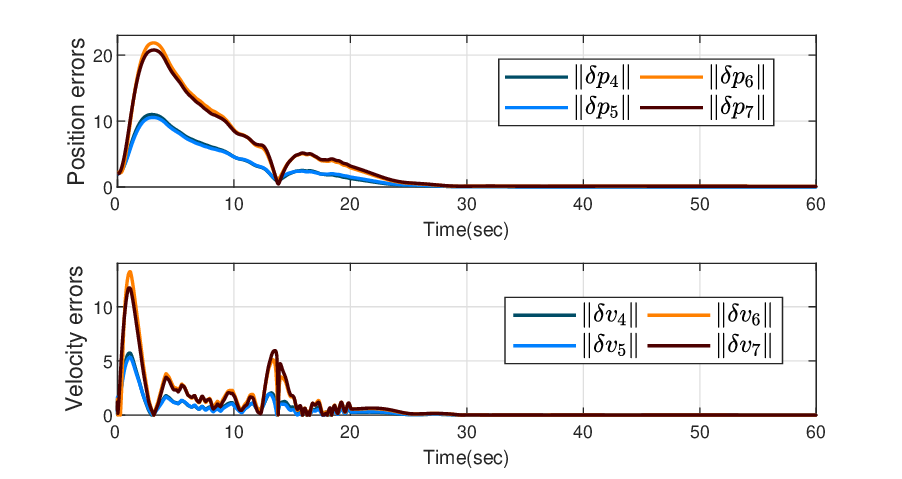}
	\caption{Position errors $\delta_{p_i}$ and velocity errors $\delta_{v_i}$ of all the agents in 2D.}
	\label{control}
\end{figure}
\begin{figure}[htpb]
	\centering
	\includegraphics[scale=0.51]{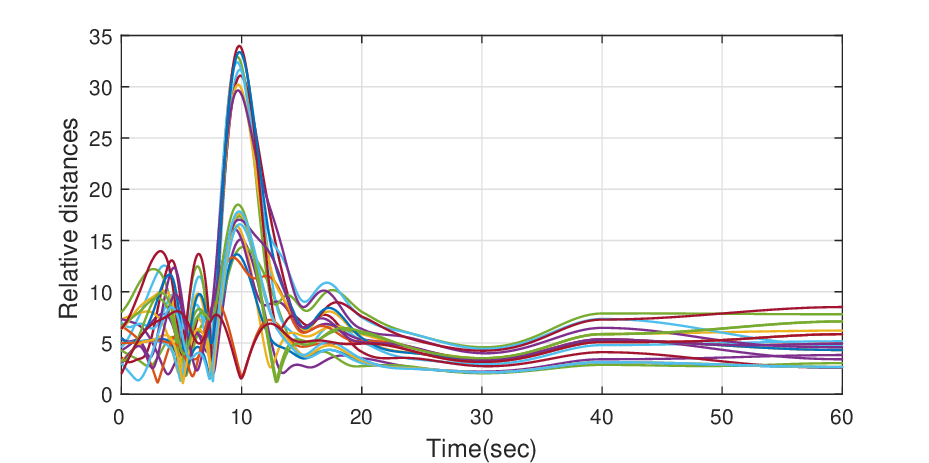}
	\caption{Distances among the agents and the target in 2D.}
	\label{dist}
\end{figure}

The trajectories of all the agents and the target are shown in Fig.\ref{trajectory}. Aside from rotation and scaling maneuvers as in \citep{yang_distributed_2021}, it can be seen that the shape deformation is achieved. All the agents enclose the target as a whole with varying radii. 
Fig.\ref{est} shows the convergence of the estimation errors of all the distances among the followers and their neighbors. Fig.\ref{control} shows that the control errors also converge to zero, which verifies the effectiveness of the robust estimation-based control method in 2D. Fig.\ref{dist} shows that the inter-agent distances are all larger than zero, which means the collision avoidance is guaranteed by choosing the initial states to satisfy \eqref{avoidance condition}.

\section{Conclusion}
In this paper, the bearing-based cooperative target entrapping control of multiple uncertain agents with arbitrary maneuvers is solved. 
A leader-follower structure is used, where the leaders move with the predesigned trajectories, and the followers are steered by a robust estimation-based control method. 
With proper design of the leaders' trajectories and a geometric configuration, an affine matrix is determined so that the persistently exciting conditions of the inter-agent relative bearings  can be satisfied. The asymptotic convergence of the estimation error and control error is proved using Filipov properties and cascaded system theories.
A sufficient condition for inter-agent collision avoidance is proposed. In the future, the targets with unknown velocities will be considered.

\bibliographystyle{unsrt}
\bibliography{templateArxiv.bib}

 

\end{document}